\documentclass[aps,pre,reprint,showpacs,superscriptaddress]{revtex4-2}
\usepackage[colorlinks,linkcolor=red,anchorcolor=red,citecolor=blue,urlcolor=blue]{hyperref}
\usepackage{amsmath,amsfonts}
\usepackage{graphicx}% Include figure files
\usepackage{dcolumn}% Align table columns on decimal point
\usepackage{float}
\usepackage{enumitem}
\allowdisplaybreaks[3]

\begin{document}

\title{Correlation between transition probability and network structure in epidemic model}

\author{Chao-Ran Cai}\email{ccr@nwu.edu.cn}
\address{School of Physics, Northwest University, Xi'an 710127, China}
\address{Shaanxi Key Laboratory for Theoretical Physics Frontiers, Xi'an 710127, China}
\author{Dong-Qian Cai}
\address{School of Physics, Northwest University, Xi'an 710127, China}

\date{\today}
\begin{abstract}
%In continuous-time dynamics, events are conceptualized as occurring continuously at a specific rate, while 
In discrete-time dynamics, 
%events are coarse-grained to occur simultaneously at a time interval with a certain transition probability.
it is frequently assumed that the transition probabilities (e.g., the recovery probability) are independent of the network structure. However, there is a lack of empirical evidence to support this claim in large time intervals.
This paper presents the nonlinear relations between the rates (in continuous-time dynamics) and probabilities of the susceptible-infected-susceptible model on annealed and static networks.
It is shown that the transition probabilities are affected not only by the rates and the time interval, but also by the network structure.
The correctness of the nonlinear relations on networks is verified based on theoretical calculation and Monte Carlo simulation.
\end{abstract}

%\pacs{}
\maketitle

\section{INTRODUCTION}

Dynamical model as a common tool in the natural and social sciences is employed to explore the characteristics of systems, predict their behavior, and analyze empirical data~\cite{Maynardsmith1974,Hofbauer1998,Barrat,fu2013propagation}.
Two distinct approaches to modelling~\cite{zhang2022nonlinear,https://doi.org/10.1002/andp.202400078}, namely continuous-time and discrete-time dynamics, have been developed and each has its own set of advantages.
Continuous-time versions consider asynchronous updates with variable time steps~\cite{Barrat}, where events (such as the infection of susceptible individuals in disease dynamics~\cite{pastor2015epidemic}) occur at certain rates.
These versions encompass the entire process of system evolution, thereby facilitating the identification of physical mechanisms and comprehension of associated phenomena.
Discrete-time versions are performed with synchronous updates at fixed time steps~\cite{Gmez_2010,cai2014effective}, wherein events occur with the state transition probabilities rather than rates, which is more straightforward to implement on a computer~\cite{PhysRevE.110.014302}.
The manner in which data is collected in the real world, including the actual epidemic processes~\cite{doi:10.1089/hs.2019.0022,PhysRevX.10.041055}, provides a rationale for the use of discrete-time version.
Integrating these two approaches can be challenging, particularly when maintaining coherence across different time intervals~\cite{https://doi.org/10.1002/andp.202400078}.

The discrete-time dynamic models provide a coarse-grained evolution process by neglecting the number of state transitions within a given time interval~\cite{chang2021analytical}, which allows for the emergence of complex behaviours such as periodic trajectories and chaos~\cite{allen1994some,vilone2011chaos,Ostilli_2013,WANG2019292}.
Some researchers have conducted comparative studies on continuous- and discrete-time dynamics~\cite{Ostilli_2013,fennell2016limitations,PhysRevE.110.014302}, revealing the equivalence of the two dynamics when the time interval tends to zero and the limitation of linear discretization of the continuous-time dynamics.
In the susceptible-infected-susceptible (SIS) model, Chang and Cai clearly indicated that the parameter mapping region between the two versions of dynamics was incomplete~\cite{chang2021analytical}, based on an anomaly that susceptible individuals exhibited a greater number of infected neighbours than infected individuals. 
In order to circumvent this drawback, Zhang and co-workers proposed a nonlinear relation between rates and probabilities~\cite{zhang2022nonlinear}, which was achieved through an analysis of the dynamic coupling effect of multiple events.
However, the homogeneity of nodes is not a valid assumption in network dynamics, as the network structure, including the degree distribution~\cite{PhysRevLett.86.3200} and dynamic correlation~\cite{PhysRevLett.116.258301}, can give rise to significant discrepancies between nodes.
Therefore, it remains uncertain whether this relation can be effectively applied to network dynamics.

The SIS model is a fundamental model for the study of transmission phenomena~\cite{pastor2015epidemic,DEARRUDA20181}. It divides the population into two categories: susceptible (S) and infected (I). The link between S and I spreads the disease with a probability $\beta'$ (or rate $\beta$), and the infected individual becomes susceptible again with a probability $\mu'$ (or rate $\mu$).
However, in the context of epidemiological processes on networks, it is common practice to assume that the infection probability ($\beta'$) and the recovery probability ($\mu'$) are identical for each individual. This is despite the fact that the structural properties of the network, such as the degree distribution, give rise to significant individual heterogeneity.
The relation between transition probabilities and network structure remains unclear.

In recent years, there have been a notable advancement in the theoretical development of the SIS model on networks~\cite{pastor2015epidemic}, largely due to the emergence of several innovative physical concepts, including heterogeneous mean-field theory~\cite{PhysRevLett.86.3200,gomez2011nonperturbative}, quenched mean-field theory~\cite{Gmez_2010,qmf_1,qmf_2}, effective degree approach~\cite{lindquist2011effective,Gleeson_1,cai2014effective}, and pair quenched mean-field theory~\cite{Mata_2013,Matamalaseaau4212,PhysRevE.102.012313}.
The same physical idea is invariably accompanied by both continuous- and discrete-time versions of the theoretical approach. 
However, the equivalence of these two versions remains undetermined, particularly when they appear to be markedly disparate.

In this paper, we examine the equivalence of continuous- and discrete-time SIS model on networks for heterogeneous mean-field theory and effective degree approach.
These expand the application of nonlinear mapping relation in Ref.~\cite{zhang2022nonlinear} to annealed and static networks.
Moreover, it is observed that the nonlinear relation between the rate and probability of each node is distinct, with the specific relation contingent upon the network structure.

The rest of this paper is organized as follows. In Sec.~\ref{model}, we undertake a review of the linear and nonlinear mapping relations between rates and probabilities.
In Sec.~\ref{Theory}, we present the nonlinear relations between rates and probabilities on the annealed networks and the static networks. And then, we verify them using the heterogeneous mean-field theory, the effective degree approach, and the Monte Carlo simulations.
In Sec.~\ref{conclusion}, we summarize our results.

\section{Review of mapping between rates and probabilities for SIS model} \label{model}
\subsection{Linear mapping relation} 
For continuous-time SIS dynamics, susceptible individuals become infected through each of their infected neighbours at a rate $\beta$ per infected neighbor, while infected individuals recover at a rate $\mu$. 
The ``rate" here refers to the number of transitions (events) that occur per unit time and is an instantaneous quantity.
For discrete-time SIS dynamics, time is no longer viewed as a continuous variable but as a discrete variable, which advances in time intervals of length $\Delta t$.
In a single time interval, susceptible individuals become infected through each of infected neighbors with probability $\beta'$ per infected neighbor, while infected individuals recover with probability $\mu'$.

The assumption of linear mapping is that the probability is equal to the product of the corresponding rate and time interval~\cite{gomez2011nonperturbative},
\begin{equation}
\begin{aligned} \label{e1}
\beta' &= \beta\Delta t,   \\
\mu' &= \mu\Delta t.  
\end{aligned}
\end{equation}
This mapping relation is generally true when $\Delta t\to0$.

\subsection{Nonlinear mapping relation} 
When the time interval $\Delta t$ is of a considerable length, individuals may undergo a number of state transitions within that interval.
These transformations do not occur in isolation; rather, they occur in a sequential and alternating manner, thereby creating coupling effects~\cite{zhang2022nonlinear}.
For the SIS model, the nonlinear mapping relation between probabilities and rates is written as follows~\cite{zhang2022nonlinear}
\begin{equation}
\begin{aligned} \label{e2}
F'(t,\Delta t)&= \frac{\langle F \rangle}{\langle F \rangle + \mu} \left[ 1 - e^{-(\langle F \rangle + \mu)\Delta t} \right],   \\
\mu' &= \frac{\mu}{\langle F \rangle + \mu} \left[ 1 - e^{-(\langle F \rangle + \mu)\Delta t} \right],
\end{aligned}
\end{equation}
%$\langle I \rangle=\frac{1}{\Delta t}\int_t^{t +\Delta t}I(\tau)d\tau$ is the average number of infected individuals within the time interval $[t,t +\Delta t]$ and
where $F'(t,\Delta t)=1 - \prod_{j\in\Omega}[1-\beta'\Theta_j(t)]$ is the probability that a susceptible individual at time $t$ transforms into infected individual at time $t+\Delta t$ and $\langle F \rangle=\frac{1}{\Delta t}\int_t^{t +\Delta t}\beta\sum_{j\in\Omega}\Theta_j(\tau)d\tau$ is the average infection rate within the time interval $[t,t +\Delta t]$.
Here, $\Omega$ is the set of neighbors of the selected individual and $\Theta_j$ is the probability that the neighbor $j$ is an infected individual. 

Once the population has reached its steady state, we have $\Theta_j(t)=\Theta_j$, Eq.~\eqref{e2} can be rewritten as
\begin{equation}
\begin{aligned} \label{e3}
 1 - \prod_{j\in\Omega}(1-\beta'\Theta_j) &= \frac{\beta \sum_{j\in\Omega}\Theta_j\left[ 1 - e^{-\left(\beta \sum_{j\in\Omega}\Theta_j + \mu\right)\Delta t} \right]}{\beta \sum_{j\in\Omega}\Theta_j + \mu} ,   \\
\mu' &= \frac{\mu\left[ 1 - e^{-\left(\beta \sum_{j\in\Omega}\Theta_j + \mu\right)\Delta t} \right]}{\beta \sum_{j\in\Omega}\Theta_j + \mu}.
\end{aligned}
\end{equation}
For fully connected networks, $\Theta_j=I/N$, the relation of Eq.~\eqref{e3} has been observed to function optimally for large $\Delta t$.

\section{Nonlinear mapping relations for annealed and static networks}\label{Theory}
\subsection{Annealed networks}\label{annealed network}
Annealed networks are defined as a situation where the network is constantly rewired at a timescale that is significantly shorter (faster) than the characteristic timescale of the disease dynamics, which may be considered the limit of rapid change.
The annealed network structure is characterised by a degree distribution $P(k)$, as the neighbours of individuals can be considered to be randomly distributed.

For the continuous-time SIS model on annealed networks, the heterogeneous mean-field theory is capable of providing a robust theoretical prediction~\cite{PhysRevLett.86.3200}.
Considering that the rates are related to the network structure, the dynamical equations are as follows:
\begin{equation}\label{e4}
\frac{dI_k(t)}{dt} = \beta_k k S_k(t) \Theta(t) - \mu_k I_k(t).  
\end{equation}
Here, $\beta_k$ ($\mu_k$) is the infection (recovery) rate for individuals of degree $k$, $I_k$ ($S_k$) represents the number of infected (susceptible) individuals with degree $k$, $\Theta=\frac{1}{N \langle k \rangle}\sum_{k}k I_k$ represents the probability that any individual is connected to an infected individual, and $I_k+S_k=NP(k)$.
For the discrete-time version, the difference equation can be written as
\begin{equation}\label{e5}
I_{k}(t+\Delta t) =I_{k}(t)(1-\mu'_k)+ S_{k}(t)\left[1 - \left(1-\beta'_k\Theta\right)^k\right],  
\end{equation}
where $\beta'_k$ ($\mu'_k$) is the infection (recovery) probability for individuals of degree $k$.

If one assumes that state transitions are independent of the network structure, then it can be concluded that 
\begin{subequations}
\begin{align}
\beta_1= \beta_2= \cdots= \beta,\ \ &\mu_1= \mu_2= \cdots= \mu, \label{6a} \\
\beta'_1= \beta'_2= \cdots= \beta',\ \ &\mu'_1= \mu'_2= \cdots= \mu'. \label{6b}
\end{align}
\end{subequations}
To establish the one-to-one mapping relation between rates and probabilities, the nonlinear mapping relation of Eq.~\eqref{e3} can be approximately simplified
\begin{equation}
\begin{aligned} \label{e7}
1 - \left(1-\beta'\Theta\right)^{\langle k \rangle} &= \frac{\beta \langle k \rangle\Theta}{\beta \langle k \rangle\Theta + \mu}\left[ 1 - e^{-\left(\beta \langle k \rangle\Theta + \mu\right)\Delta t} \right], \\
\mu' &= \frac{\mu}{\beta \langle k \rangle\Theta + \mu}\left[ 1 - e^{-\left(\beta \langle k \rangle\Theta + \mu\right)\Delta t} \right].
\end{aligned}
\end{equation}

Eliminating the restriction of Eq.~\eqref{6a} and Eq.~\eqref{6b}, the nonlinear mapping of Eq.~\eqref{e3} can be simplified as
\begin{equation}
\begin{aligned} \label{e8}
1 - \left(1-\beta'_k\Theta\right)^{k} &= \frac{\beta_k k\Theta}{\beta_k k\Theta + \mu_k}\left[ 1 - e^{-\left(\beta_k k\Theta + \mu_k\right)\Delta t} \right], \\
\mu'_k &= \frac{\mu_k}{\beta_k k\Theta + \mu_k}\left[ 1 - e^{-\left(\beta_k k\Theta + \mu_k\right)\Delta t} \right].
\end{aligned}
\end{equation}
In the event that a pair of rates $(\beta,\mu)$ is provided, Eq.~\eqref{6a} and Eq.~\eqref{e8} can be used to obtain
\begin{equation}\label{e9}
\frac{d\mu'_k}{dk}\Big|_{(\beta,\mu)}<0, \ \ \frac{d}{dk}\left[1 - \left(1-\beta'_k\Theta\right)^{k}\right]\Big|_{(\beta,\mu)}>0.
\end{equation}
As illustrated in Eq.~\eqref{e9}, the recovery probability of an individual decreases with the increase of degree $k$, whereas the total infection probability increases with the increase of degree $k$.
Similarly, if a pair of probabilities $(\beta',\mu')$ is provided, Eq.~\eqref{6b} and Eq.~\eqref{e8} can be used to obtain 
\begin{equation}\label{e10}
\frac{d\mu_k}{dk}\Big|_{(\beta',\mu')}>0, \ \ \frac{d\left(\beta_kk\Theta\right)}{dk}\Big|_{(\beta',\mu')}>0.
\end{equation}
Equation~\eqref{e10} indicates that both the recovery rate and the total infection rate increase with the degree $k$.
Moreover, Eq.~\eqref{e8} can also derive the inequality
\begin{equation}\label{e11}
\mu'_k<\left(1-\beta'_k\widetilde{\Theta}\right)^{k},
\end{equation}
which constrains the applicability of the mapping relations between probabilities and rates.
Here, $\widetilde{\Theta}=\max\{\Theta(t=0),\Theta(t=\infty)\}$.
Any pair of probabilities that do not satisfy Eq.~\eqref{e11} is not accompanied by corresponding rates.
Meanwhile, Eq.~\eqref{e11} shows that at least one of the probabilities $\mu'_k$ and $\beta'_k$ is very small for a large $k$.
The derivation of Eqs.~\eqref{e9}-\eqref{e11} is presented in the Appendix~\ref{app}.

In Fig.~\ref{fig1}, we assess the consistency of the discrete-time SIS model with the continuous-time version by employing various mapping relations between probabilities and rates over a large time interval.
The initial step involve an examination of the prevalence $\rho$ of various infection rates on homogeneous and heterogeneous networks, where $\rho=\rho(t=\infty)=(1/N)\sum_{k}I_k(t=\infty)$.
The results yielded by the nonlinear relation of Eq.~\eqref{e7} are consistent with the data of continuous-time dynamics in Fig.~\ref{fig1}(a), but a minor discrepancy is observed in Fig.~\ref{fig1}(b).
It is encouraging to note that Eq.~\eqref{e8} performs well in both Fig.~\ref{fig1}(a) and (b), where it is a nonlinear relation without the constraint of one-to-one mapping.

\begin{figure}
\centering
\includegraphics[width=\linewidth]{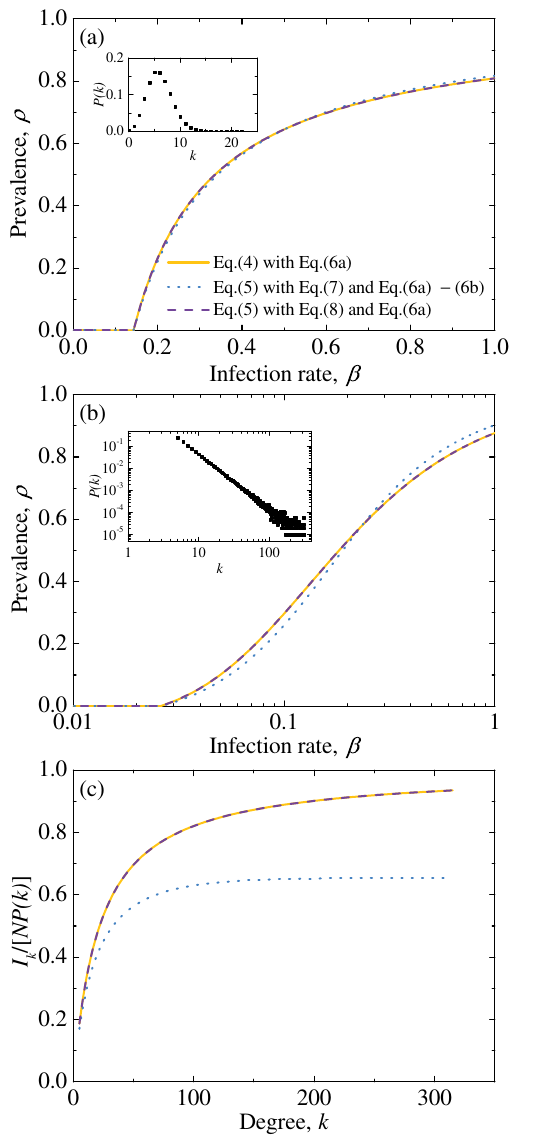}
\caption{Mapping from continuous-time SIS model to discrete-time SIS model on annealed networks. 
(a)-(b) Epidemic prevalence $\rho$ as a function of infection rate $\beta$.
(c) Distribution of infected individuals by degree $k$.
The degree distribution $P(k)$ adopts a Poisson distribution with $\langle k \rangle=6$ in (a) and a scale-free distribution with $k_{\min}=5$ and $\gamma=2.5$ in (b) and (c), as illustrated in the inset panels.
The yellow lines represent the results of continuous-time dynamics, while the other lines represent the results of discrete-time dynamics using nonlinear relations.
Parameters: $\mu = 1$, the fraction of individuals initially infected $\rho_0=0.01$, $\Delta t = 1$ in discrete-time dynamics, $\beta=0.1$ in (c), $N = 10^6$ in (a) and $N = 10^5$ in (b)-(c). The dash and dot lines are used to enhance the intuitiveness of overlapping data.}\label{fig1}
\end{figure}
\begin{figure}
\centering
\includegraphics[width=\linewidth]{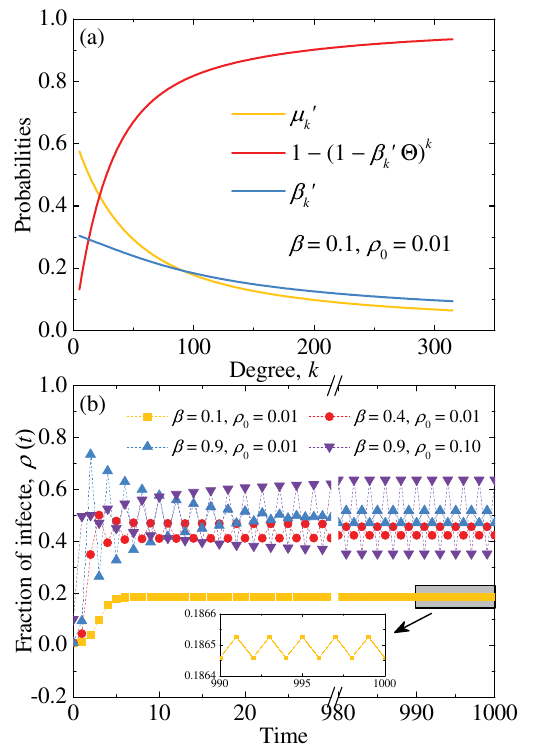}
\caption{(a) The probabilities $\mu'_k$, $1 - \left(1-\beta'_k\Theta\right)^{k}$, and $\beta'_k$ as a function of degree $k$ when the system reaches steady state. 
(b) The fraction of infected individuals as a function of the time for discrete-time SIS model on annealed networks. 
The results in (a) are obtained from Eq.~\eqref{e5} and Eq.~\eqref{e8}, and the results in (b) are obtained from Eq.~\eqref{e5} and Eq.~\eqref{e1} (the linear relations).
The degree distribution $P(k)$ is consistent with Fig.~\ref{fig1}(b). 
Parameters: $\mu = 1$, $\Delta t = 1$.}\label{fig2}
\end{figure}

To gain further insight into the sources of differences, we conduct a detailed analysis of the distribution of infected individuals on heterogeneous networks in Fig.~\ref{fig1}(c).
The results obtained from Eq.~\eqref{e7} exhibit a significant discrepancy in the region of high degrees, whereas the results yielded by Eq.~\eqref{e8} demonstrate a remarkable consistency for all $k$.
Thus, the infection and recovery probabilities are related to the network structure of the annealed network, even if those rates are set to be independent of the network.

Figure~\ref{fig2}(a) depicts the infection and recovery probabilities of varying degrees when the system reaches steady state for a specific pair of rates $(\beta=0.1,\mu = 1)$.
As the degree $k$ increases, the recovery probability declines, while the total infection probability rises. This is consistent with the prediction of inequality in Eq.~\eqref{e9}.

Finally, Fig.~\ref{fig2}(b) shows that the linear relation from Eq.~\eqref{e1} is not adequate.
In such cases, a stable trajectory of period-2 emerges instead of the steady state.
In the case where $\mu=1$ and $\Delta t = 1$, the transition probabilities obtained from Eq.~\eqref{e1} are inconsistent with the constraints of Eq.~\eqref{e11}.

\subsection{Static network}\label{static network}

A static network is defined as a network on a time scale significantly longer (slower) than the time scale characteristic of disease dynamics, it is a network with fixed edges and can also be thought of as a slowly changing boundary.
The static network structure is characterized by an adjacency matrix, which includes degree distribution and fixed edges, the latter producing dynamic correlation~\cite{PhysRevLett.116.258301,cai2020analytical,PhysRevE.103.032313}.

For the continuous SIS model on a static network without degree-degree correlations, the effective
degree approach can provide a good theoretical prediction~\cite{lindquist2011effective,Gleeson_1,cai2014effective}. 
The effective degree approach considers not only the state transitions of the individuals themselves but also those of their neighbors. 
Considering that the rates are contingent upon the network structure, the dynamical equations are as follows:
\begin{equation}
\begin{aligned} \label{e12}
\frac{d S_{si}}{dt} = & -\beta_{si} i S_{si} + \mu_{si} I_{si} + M_S\left[(i+1) S_{s-1, i+1} - i S_{si}\right] \\
& + B_S\left[(s+1) S_{s+1, i-1} - s S_{si}\right], \\
\frac{d I_{si}}{dt} = & \beta_{si} i S_{si} - \mu_{si} I_{si} + M_I\left[(i+1) I_{s-1, i+1} - i I_{si}\right] \\
& + B_I\left[(s+1) I_{s+1, i-1} - s I_{si}\right].
\end{aligned}
\end{equation}
Here, $\sum_{s+i=k}(I_{si}+S_{si})=NP(k)$. 
$\beta_{si}$ ($\mu_{si}$) is the infection (recovery) rate of an individual with $s$ susceptible neighbors and $i$ infected neighbors.
$I_{si}$ ($S_{si}$) represents the number of infected (susceptible) individuals with $s$ susceptible neighbors and $i$ infected neighbors.
The average infection and recovery rates for neighbors of susceptible (infected) individuals are represented by $B_S$ ($B_I$) and $M_S$ ($M_I$), respectively, with
\begin{equation}
\begin{aligned} \label{e13}
M_S=\frac{\sum_{k} \sum_{j+l=k} j \mu_{jl} I_{jl}}{\sum_{k} \sum_{j+l=k} l S_{jl}}, B_S=\frac{\sum_{k} \sum_{j+l=k} j \beta_{jl} l S_{jl}}{\sum_{k} \sum_{j+l=k} j S_{jl}},\\
M_I=\frac{\sum_{k} \sum_{j+l=k} l\mu_{jl} I_{jl}}{\sum_{k} \sum_{j+l=k} l I_{jl}}, B_I=\frac{\sum_{k} \sum_{j+l=k} \beta_{jl} l^{2} S_{jl}}{\sum_{k} \sum_{j+l=k} j I_{jl}}.
\end{aligned}
\end{equation}
The concept of mean field is employed when assessing the transition rates of the neighbors in Eq.~\eqref{e13}.

For the discrete-time version, combined with the effective degree Markov-chain approach in Ref.~\cite{cai2014effective} and the probabilities depending on the network structure, the difference equation can be written as
\begin{equation}\label{e14}
\begin{aligned}
S_{si}(t+\Delta t) = \sum_{j+l=s+i}&\left\{S_{jl}(t) (1-\beta'_{jl})^l G_{jl\rightarrow si} \right.\\
&\left.+I_{jl}(t) \mu'_{jl}F_{jl\rightarrow si}\right\}, \\
I_{si}(t+\Delta t) = \sum_{j+l=s+i}&\left\{S_{jl}(t) \left[1-(1-\beta'_{jl})^l\right]G_{jl\rightarrow si} \right.\\
&+ \left. I_{jl}(t) (1-\mu'_{jl})F_{jl\rightarrow si}\right\},
\end{aligned}
\end{equation}
where $\beta'_{jl}$ ($\mu'_{jl}$) is the infection (recovery) probability of an individual with $j$ susceptible neighbors and $l$ infected neighbors.
Here, the nonlinear relation expressed by Eq.~\eqref{e3} can be succinctly represented as
\begin{equation}
\begin{aligned}\label{e15}
1-(1-\beta'_{jl})^l =  \frac{\beta_{jl} l}{\beta_{jl} l + \mu_{jl}}\left[1 - e^{-(\beta_{jl} l + \mu_{jl})\Delta t}\right], \\
\mu'_{jl} =  \frac{\mu_{jl}}{\beta_{jl} l + \mu_{jl}}\left[1 - e^{-(\beta_{jl} l + \mu_{jl})\Delta t}\right].
\end{aligned}
\end{equation}
$G_{jl\rightarrow si}$ and $F_{jl\rightarrow si}$ are the probabilities of the subscript transformations for $S_{jl}$ and $I_{jl}$, respectively, and their specific forms are as follows
\begin{equation}\label{e16}
\begin{aligned}
G_{jl\rightarrow si} =  \sum_p  \bigg[& \binom{l}{p}(M'_{S})^p(1-M'_{S})^{l-p}\binom{j}{p+j-s}   \\
&   (B'_{S})^{p+j-s}(1-B'_{S})^{s-p}  \bigg],\\
F_{jl\rightarrow si} =  \sum_p  \bigg[& \binom{l}{p}(M'_{I})^p(1-M'_{I})^{l-p}\binom{j}{p+j-s}  \\
&   (B'_{I})^{p+j-s}(1-B'_{I})^{s-p} \bigg],
\end{aligned}
\end{equation}
where $p$ is the number of infected neighbors who have recovered and is constrained by the condition $\max\{0,s-j\}\leq p\leq\min\{s,l\}$.
The average infection and recovery probabilities for neighbors of susceptible (infected) individuals are represented by $B'_{S}$ ($B'_{I}$) and $M'_{S}$ ($M'_{I}$), respectively.

\begin{figure}
\centering
\includegraphics[width=\linewidth]{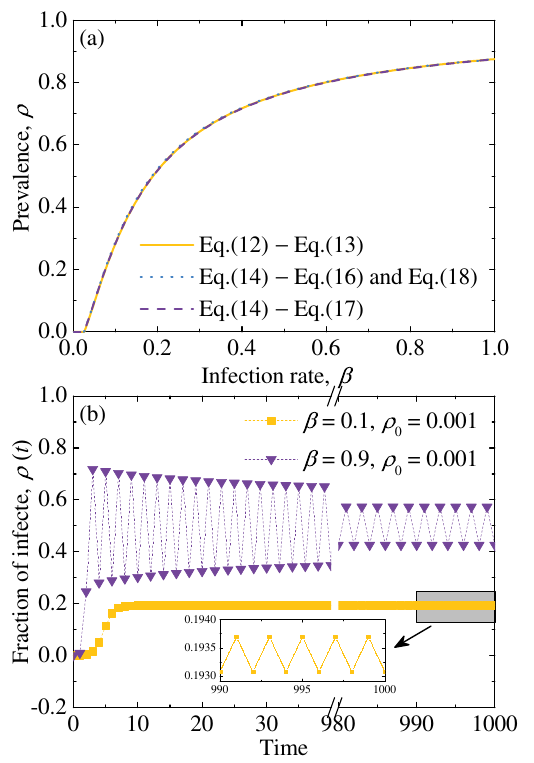}
\caption{Mapping from continuous-time SIS model to discrete-time SIS model on static networks. 
(a) Epidemic prevalence $\rho$ as a function of infection rate $\beta$.
(b) The fraction of infected individuals as a function of the time for discrete-time SIS model on static networks. The results in (b) are obtained from Eq.~\eqref{e14}, Eq.~\eqref{e16}, Eq.~\eqref{e17}, and Eq.~\eqref{e1} (the linear relations).
The static scale-free networks is generated from the uncorrelated configuration model~\cite{PhysRevE.71.027103} with power-law degree distributions $P(k)\sim k^{-\gamma}$ where $\gamma=2.5$ and $k_{\min}=5$.
Parameters: $\mu = 1$, $\rho_0=0.001$, $\Delta t = 1$ in discrete-time dynamics, $N = 10^5$. }\label{fig3}
\end{figure}

The four subscript transition probabilities can be calculated in two ways. 
One method is to combine the probabilities in Eq.~\eqref{e15} with the concept of mean field to derive 
\begin{equation}
\begin{aligned} \label{e17}
B'_S=&\frac{\sum_k \sum_{j+l=k} j S_{jl}\left[1-(1-\beta'_{jl})^{l}\right]}{\sum_k \sum_{j+l=k} j S_{jl}}, \\
M'_S=&\frac{\sum_k \sum_{j+l=k} j I_{jl}\mu'_{jl}}{\sum_k \sum_{j+l=k} l S_{jl}}, \\
B'_I=&\frac{\sum_k \sum_{j+l=k} l S_{jl}\left[1-(1-\beta'_{jl})^{l}\right]}{\sum_k \sum_{j+l=k} j I_{jl}}, \\
M'_I=&\frac{\sum_k \sum_{j+l=k} l I_{jl}\mu'_{jl}}{\sum_k \sum_{j+l=k} l I_{jl}}.
\end{aligned}
\end{equation}
The other method is to combine Eqs.~\eqref{e3} and \eqref{e13} to obtain 
\begin{equation}\label{e18}
\begin{aligned}
B'_{S} &= B_S \frac{1-e^{-(B_S+M_S)\Delta t}}{B_S+M_S}, \\
M'_{S} &= M_S \frac{1-e^{-(B_S+M_S)\Delta t}}{B_S+M_S},\\
B'_{I} &= B_I \frac{1-e^{-(B_I+M_I)\Delta t}}{B_I+M_I}, \\
M'_{I} &= M_I \frac{1-e^{-(B_I+M_I)\Delta t}}{B_I+M_I}.
\end{aligned}
\end{equation}
It is straightforward to demonstrate that Eq.~\eqref{e17} and Eq.~\eqref{e18} are equivalent as $\Delta t$ approaches zero.
Furthermore, as will be shown subsequently in Fig.~\ref{fig3}(a), they are also approximately equivalent for large $\Delta t$.
Finally, the following constraint relation between probabilities exists in the static network,
\begin{equation}
\mu'_{jl}<(1-\beta'_{jl})^l,
\end{equation}
which is derived from Eq.~\eqref{e15}.

To reinforce the comparative effect, we set the rates independently of the static network in Fig.~\ref{fig3}.
Figure~\ref{fig3}(a) serves to verify the validity of nonlinear mapping relations on static networks for large $\Delta t$.
Meanwhile, it is elucidated that the infection and recovery probabilities are contingent upon the network structure of the static network.
Figure~\ref{fig3}(b) shows a stable trajectory of period-2 again for $\mu=1$ and $\Delta t =1$, indicating the limitation of the linear relations.

\subsection{Results from Monte Carlo simulations} \label{result}

\begin{figure}
\centering
\includegraphics[width=\linewidth]{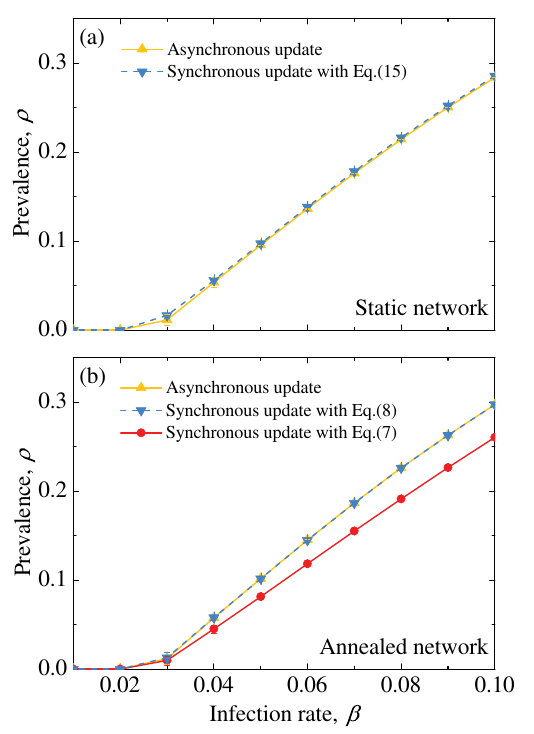}
\caption{Monte Carlo simulations for SIS model on annealed and static networks. Epidemic prevalence $\rho$ as a function of infection rate $\beta$ for different update mode and mapping relations. 
The static network and annealed network are consistent with Fig.~\ref{fig3} and Fig.~\ref{fig1}(b), respectively.
Parameters: $\mu = 1$, $N = 10^5$, $\rho_0=0.001$, $\Delta t = 1$ in synchronous update.}\label{fig4}
\end{figure}

Next, we verify the validity of the nonlinear mapping relations from Monte Carlo simulations, ensuring that no approximation processing is involved.
For the sake of simplicity, Fig.~\ref{fig4} makes the assumption that the infection rate of all edges and the recovery rate of all nodes are identical, which is consistent with the configuration depicted in Figs.~\ref{fig1}-\ref{fig3}.
Figures~\ref{fig4}(a) and \ref{fig4}(b) show the efficacy of Eq.~\eqref{e15} and Eq.~\eqref{e8} in static and annealed networks, respectively. 
This further corroborates the assertion that infection probability and recovery probability are contingent upon network structure.
More precisely, these probabilities are related to the number or expected number of infected immediate neighbors, which is represented by $l$ in Eq.~\eqref{e15} or $k\Theta$ in Eq.~\eqref{e8}.
Subsequently, we present the detailed Monte Carlo simulation procedures.

For the yellow triangles in Fig.~\ref{fig4}(a), continuous-time stochastic simulations on static networks are implemented using the optimized Gillespie algorithm~\cite{cota2017optimized}, which is an asynchronous update algorithm.
The simulation procedure works as follows:

Step 1: Construct a list $\nu$ of individuals that have been infected.

Step 2: At any time $t$, calculate the total rate $\omega(t)=\sum_{i\in I(t)}(\mu+\beta k_i)$, where $k_i$ is the degree of node $i$.
And then, Step 2a should be performed with probability $\frac{\mu I(t)}{\omega(t)}$, otherwise Step 2b should be performed with probability $1-\frac{\mu I(t)}{\omega(t)}$.

\begin{itemize}
 \item Step 2a: An infected individual $i$ is chosen with equal probability from $\nu$ and cured.
 \item Step 2b: An infected individual $i$ is chosen from $\nu$ at random and accepted with probability $k_i/k_{\max}$, which is repeated until one choice is accepted.
And then, a neighbor of $i$ is chosen randomly. If the neighbor is susceptible, it becomes infected.
\end{itemize}

Step 3: The time is incremented by $dt = \frac{1}{\omega(t)}$. The list $\nu$ is updated.

Step 4: Repeat Steps 2-3 until the predetermined time period is reached.

For the blue inverse triangles in Fig.~\ref{fig4}(a), discrete-time stochastic simulations on static networks are performed using the synchronous update algorithm mentioned in Refs.~\cite{cai2014effective,chang2021analytical,PhysRevE.110.014302}.
Note that $\beta'_{jl}$ and $\mu'_{jl}$ mentioned later must be calculated according to Eq.~\eqref{e15}.
The simulation procedure works as follows:

Step 1: At any time step $t$, calculate each individual's transition probability. 
The infection probability for a susceptible individual is $1-(1-\beta'_{jl})^l$.
The recovery probability for a infected individual is $\mu'_{jl}$.
Here, the subscripts $j$ and $l$ represent that the individual in question has $j$ susceptible neighbors and $l$ infected neighbors.

Step 2: At time step $t + \Delta t$, all individuals update their states in a synchronous way according to the probabilities of Step 1.

Step 3: Repeat Steps 1-2 until the predetermined time period is reached.

For the yellow triangles in Fig.~\ref{fig4}(b), continuous-time stochastic simulations on annealed networks are implemented using the asynchronous update algorithm mentioned in Ref.~\cite{PhysRevLett.116.258301}.
The simulation procedure works as follows:

Step 1: Construct a list $\nu$ of individuals that have been infected.

Step 2: At any time $t$, calculate the selected rate of each infected individual $i$, $\eta_i(t)=\mu+\beta k_i$, where $k_i$ is the degree of node $i$.
The total rate is $\omega(t)=\sum_{i\in I(t)}\eta_i(t)$.
The infected individual $j$ who is selected is sampled with a probability proportional to $\eta_j(t)$.
And then, Step 2a should be performed with probability $\frac{\mu}{\eta_j(t)}$, otherwise Step 2b should be performed with probability $1-\frac{\mu}{\eta_j(t)}$.

\begin{itemize}
\item Step 2a: The selected individual $j$ is cured.
\item Step 2b: Another individual $j'$ who will be contacted by $j$ is selected with probability proportional to its degree.
If the individual $j'$ is susceptible, it becomes infected.
\end{itemize}

Step 3: The time is incremented by $dt = \frac{1}{\omega(t)}$. The list $\nu$ is updated.

Step 4: Repeat Steps 2-3 until the predetermined time period is reached.

For the blue inverse triangles and the red circles in Fig.~\ref{fig4}(b), discrete-time stochastic simulations on annealed networks are performed using the following  synchronous update algorithm.
The simulation procedure works as follows:

Step 1: At any time step $t$, calculate each individual's transition probability. 
For any given individual $j$, the neighbors that it contacts are selected, and the probability of each selection is proportional to the degree of the selected individual.
In the event that individual $j$ is susceptible, the infection probability is $1-(1-\beta'_{k})^{k_{\inf}}$, where $k_{\inf}$ is the number of infected neighbors.
Conversely, if individual $j$ is infected, the recovery probability is $\mu'_{k}$.

Step 2: At time step $t + \Delta t$, all individuals update their states in a synchronous way according to the probabilities of Step 1.

Step 3: Repeat Steps 1-2 until the predetermined time period is reached.

Note that $\beta'_{k}$ and $\mu'_{k}$ are calculated according to Eq.~\eqref{e8} for blue inverse triangles or Eq.~\eqref{e7} for red circles.

\section{conclusion}\label{conclusion}
In conclusion, we investigate the nonlinear mapping relations between probabilities and the rates of SIS model on annealed and static networks, with particular attention to the large time intervals and the heterogeneity of degree distribution.
The nonlinear mapping relations consider the coupling effect of multiple events occurring within a time interval to address the information loss resulting from discretization, which is the phenomenon of several changes of state.
The equivalence of continuous- and discrete-time SIS model on networks is verified through the use of nonlinear relations in three aspects: the heterogeneous mean-field theory, the effective degree approach, and the Monte Carlo simulations.
The nonlinear relations, as described by Eq.~\eqref{e8} in annealed networks and Eq.~\eqref{e15} in static networks, suggests a natural correlation between the state transition probabilities and the network structure.

In addition, quantitative criteria are provided to ascertain the applicable range of nonlinear relations. 
To test a set of parameters that do not satisfy the aforementioned quantitative criteria, we employ linear mapping relations, which finds that a stable trajectory of period-2, rather than the steady state.

It is anticipated that this methodology will be applicable to more realistic temporal networks~\cite{perra2012activity,PhysRevResearch.6.L022017} and higher-order networks~\cite{Lin2024,PhysRevLett.132.077401} in the future. 
This research line contribute to a more comprehensive understanding of the dynamic behavior of complex networks.

\begin{acknowledgments}
    This work was supported by the Shaanxi Fundamental Science Research Project for Mathematics and Physics (Grant No. 22JSQ003).
\end{acknowledgments}

\appendix 
\section{Derivation of Eq.~\eqref{e9}-\eqref{e11}}\label{app}
When specific rates $(\beta,\mu)$ is employed, substituting Eq.~\eqref{6a} and $x=(\beta k\Theta + \mu)\Delta t$ into Eq.~\eqref{e8} yields 
\begin{equation}
\begin{aligned} \label{a1}
\mu'_k &=\mu\Delta t(1 - e^{-x})x^{-1}, \\
1 - \left(1-\beta'_k\Theta\right)^{k}+\mu'_k &=1-e^{-x}.
\end{aligned}
\end{equation}
Therefore, one can get the following relations
\begin{equation}
\begin{aligned} \label{a2}
\frac{d\mu'_k}{dk}=-\beta\mu \Theta\left(1-e^{-x}-xe^{-x}\right)\left(\frac{\Delta t}{x}\right)^2<0, \\
\frac{d}{dk}\left[1 - \left(1-\beta'_k\Theta\right)^{k}\right]=\beta \Theta\Delta t\left(e^{-x}-\frac{d\mu'_k}{dx}\right)>0.
\end{aligned}
\end{equation}
Here, the inequality $1-xe^{-x}>e^{-x}$ is employed for the analysis of Eq.~\eqref{a2}.

For the case of using specific probabilities $(\beta',\mu')$, substituting Eq.~\eqref{6b} and $y=1 - \left(1-\beta'\Theta\right)^{k}$ into Eq.~\eqref{e8} yields
\begin{equation}
\begin{aligned}\label{a3} 
\beta_k k\Theta= -\frac{\ln\left(1-y-\mu'\right)}{y+\mu'}\frac{y}{\Delta t} \\
\mu_k= -\frac{\ln\left(1-y-\mu'\right)}{y+\mu'}\frac{\mu'}{\Delta t}
\end{aligned}
\end{equation}
Notice that the term $\ln\left(1-y-\mu'\right)$ is less than zero. Let $z=1-y-\mu'$, get
\begin{equation}
\begin{aligned} \label{a4}
\frac{d(\beta_kk\Theta)}{dk}=\frac{dy}{dk}\frac{\mu'}{\Delta t(1-z)^2}\left[\left(\frac{1}{z}-1\right)\frac{y}{\mu'}-\ln z\right]>0, \\
\frac{d\mu_k}{dk}=\frac{dy}{dk}\frac{\mu'}{\Delta t(1-z)^2}\left(\frac{1}{z}-1+\ln z\right)>0.
\end{aligned}
\end{equation}
Here, the inequalities $\ln z>1-1/z$ and $dy/dk>0$ is used for the analysis of Eq.~\eqref{a4}.

According to Eq.~\eqref{a1}, we have
\begin{equation}\label{a5}
\mu'_k=\left(1-\beta'_k\Theta\right)^{k}- e^{-x}<\left(1-\beta'_k\Theta\right)^{k}, 
\end{equation}
From the monotonicity of $\Theta$, it can be inferred that
\begin{equation}\label{a6}
\left(1-\beta'_k\Theta\right)^{k}\geq\left(1-\beta'_k\widetilde{\Theta}\right)^{k}, 
\end{equation}
where $\widetilde{\Theta}=\max\{\Theta(t=0),\Theta(t=\infty)\}$. 
Combining Eq.~\eqref{a5} and Eq.~\eqref{a6}, gives Eq.~\eqref{e11}.

\bibliography{ref1}
\end{document}